\documentstyle[aps,preprint]{revtex}

\newcommand{\brho}{\mbox{\boldmath $\rho$}}
\newcommand{\bmu}{\mbox{\boldmath $\mu$}}
\newcommand{\bta}{\mbox{\boldmath $\eta$}}
\begin{document}
\tolerance=100000
\author{ Zafar Ahmed \\
Nuclear Physics Division, Bhabha Atomic Research Centre \\
Trombay, Bombay 400 085, India \\ zahmed@apsara.barc.ernet.in }
\title
{Pseudo-reality and pseudo-adjointness of Hamiltonians}
\date{\today}
\maketitle
\begin{abstract}
We define pseudo-reality and pseudo-adjointness of a Hamiltonian, $H$, as
$\brho H \brho^{-1}=H^\ast$ and $\bmu H \bmu^{-1}=H^\prime$, respectively.
We prove that the former yields the {\it necessary} condition for
spectrum to be real whereas the latter helps in fixing a definition for
inner-product of the eigenstates. Here we separate out adjointness of an
operator from its Hermitian-adjointness. It turns out that a Hamiltonian
possessing real spectrum is first pseudo-real, further it could be
Hermitian, PT-symmetric or pseudo-Hermitian.
\end{abstract}
\vspace {.2 in}
\par Last few years have witnessed a new scope for even a non-Hermitian
Hamiltonians to possess real spectrum. It has been found that when a
Hamiltonian, $H$, is invariant under the joint action of parity $(P :
x\rightarrow -x)$ and time-reversal ($T : i\rightarrow -i)$ i.e. $[PT,H]=0$,
there may arise
surprisingly two situations. One, when $PT$ and $H$ admit common eigenstates
and two, when they do not do so. In the former situation one can show that
the eigenvalues will be real and in the latter the eigenvalues are conjectured
to be complex-conjugate pairs and PT-symmetry is said to be spontaneously
broken. One can indeed not tell whether a $PT$-symmetric
potential has real or complex (conjugate pairs of) energy-eigenvalues, until
the wavefunctions are analyzed. This intriguing feature has inspired the
pursuit [1-6] of both analytically and numerically solved models of PT-symmetric
potentials. Thus, PT-symmetry of a Hamiltonian could at most be a {\it
necessary} and not the sufficient condition for the reality of eigenvalues.
\par Further, the phenomenon of real eigenvalues of non-Hermitian Hamiltonians
has been found to be connected with the already known concept of pseudo-Hermiticity.
A Hamiltonian is pseudo-Hermitian [7-16] if
\begin{equation}
\bta H \bta^{-1} =H^\dagger.
\end{equation}
It has also been known that
\begin{equation}
(E^\ast_m-E_n)\Psi^\dagger_m \bta \Psi_n=0,~~~
N_{\bta,n}=\Psi^\dagger_n \bta \Psi_n.
\end{equation}
Here we propose to choose matrix notations for a subtle reason that these
notations have a separate and explicit sign for the adjoint (transpose) operation.
The sign ,$^\dagger$, jointly denotes complex-conjugation, $^\ast$, and
transpose (adjoint), $^\prime$, of the operators or vectors. Notice that (2)
merely asserts two important features of the eigenstates (i) : If eigenvalues
are real and distinct, the eigenstates will be $\bta$-orthogonal  as $\Psi^
\dagger_m \bta \Psi_n= \epsilon \delta_{m,n}$. (ii) : Complex eigenvalues will
have zero pseudo-norm i.e., $N_{\bta,n}=0$. Here, we must bring home the fact
that {\it the concept of pseudo-Hermiticity as such does not yield an explicit proof for
the reality of eigenvalues (even under any further condition), it can only support
real eigenvalues indirectly (see Eq. (2)) }. This shortcoming of pseudo-Hermiticity
which has gone un-noticed both recently and initially [7-16] motivates the
present work.
\par Several PT-symmetric potentials having real spectrum have been found to be
parity-pseudo-Hermitian where $(\bta= P)$[8]. Several complex potentials which
are both PT-symmetric and non-PT-symmetric have been found to be pseudo-Hermitian
when $\bta=e^{-\theta p_x}$ [9]. This operator affects an imaginary shift in the
co-ordinate i.e. : $ \bta x \bta^{-1}=x+i\theta.$ Several other Hamiltonians
of both the types have been reported [10] to be pseudo-Hermitian under $\bta=e^
{\phi(x)}$ : a gauge-like transformation. It has been proved that if a non-Hermitian
operator possesses real eigenvalues then there exists (one can find) a metric of the types
$\bta=OO^\dagger$ [8], or $(OO^\dagger)^{-1}$ [11] under which the Hamiltonian
is pseudo-Hermitian. Next, following matrix algebra it has been stated and
proved [24] that if a matrix-Hamiltonian has real eigenvalues and a diagonalizing
matrix $D$ then it is pseudo-Hermitian under $\bta_+=(DD^\dagger)^{-1}$ and
{\it vice versa}.
\par Pseudoanti-Hermiticity [9] and a recipe [12] for construction
of pseudo-Hermitian potentials have also been discussed.
Clearly, without knowing a metric one can not invoke pseudo-Hermiticity.
One can find at least one metric, $\bta_+$, as stated above.
It is also, known that a Hamiltonian could be pseudo-Hermitian under several metrics.
These metrics would further help in bringing out the symmetry of the Hamiltonian
as $[H,\bta_i\bta_j^{-1}]$ [8,23,24]. These
metrics may be real, complex, Hermitian, non-Hermitian, unitary, proper
($\det (\bta)=1$), involutary ($\bta^2={\bf 1}$) and {\it secular} etc.. When
a metric does not depend upon the parameters of the Hamiltonian, we call it
{\it secular} [16].
\par At this stage of the developments, we find that the adjointness of a
Hamiltonian has not been taken in to account when we discuss the PT-symmetry
or pseudo-Hermiticity of a Hamiltonian. As a result,
we find that a potential despite being both PT-symmetric and pseudo-Hermitian
and possessing real spectrum does not satisfy (e.g. [10]) the PT-orthogonality
(PT-inner-product) [3].
\begin{equation}
(E^\ast_m-E_n) \Psi^{PT}_m \Psi_n =0,~~~N_{PT,n}=\Psi^{PT~^\prime}_n \Psi_n.
\end{equation}
It, however, satisfies $\bta$-pseudo-orthogonality condition [2].
This is as though PT-symmetry is not enough to ensure orthogonality of eigenstates.
A special analysis has been carried out [15] to uphold PT-symmetry in this regard,
eventually it yielded a condition more akin to (2). Moreover, as mentioned above
the concept of pseudo-Hemiticity at best does not contradict the occurrence of the
real eigenvalues nevertheless it does not provide a proof for it. This is achived here
in the present work by introducing the concept of pseudo-reality of Hamiltonians.
\par In this letter, we introduce the concept of pseudo-reality and
pseudo-adjointness of a Hamiltonian by proposing to separate out adjointness
of an operator from the Hermitian-adjointness, a subtle point which has been
missed out in the developments described above.
\par Let us first discuss the adjointness of an operator.
We propose to use $^\prime$ sign for adjoint and transpose if the Hamiltonian
is in differential and matrix form, respectively. The adjoint of a differential
operator $A$ denoted as $A^\prime$ is defined as [17]
\begin{equation}
u. A v-v.A^\prime u={dW(u,v) \over dx},
\end{equation}
i.e. the right hand side is an exact differential and $W$ is called bilinear
concomitant [17].
The functions $u,v$ are two arbitrary vectors form a vector space.
Here the dot denotes simple multiplication. Subsequently, we have
\begin{equation}
\left ( {d^n \over dx^n} \right)^\prime =(-1)^n {d^n \over dx^n}, ~n=1,2,...
\end{equation}
Thus for the quantum mechanical operators : position, momentum and kinetic
energy, we have
\begin{equation}
(x)^\prime=x,~~ (p_x)^\prime =-p_x, ~~\mbox{and}~~ (K)^\prime=K.
\end{equation}
Thus, Hamiltonians of the type ${p_x}^2/(2m)+V(x)$ are
self-adjoint, i.e. $H=H^\prime$. Usually, we use the concept of Hermitian-adjointness
in quantum mechanics, i.e.
\begin{equation}
(p_x)^\dagger=\left (-i\hbar {d \over dx} \right )^{\prime~\ast}=p_x,
\end{equation}
and call an operator  $A\equiv  p_x$, $K$ and $x$ to be self-(Hermitian)-adjoint
by also noticing that
$\langle A\Psi|\Psi\rangle=\langle \Psi|A^\dagger \Psi \rangle$ [17,18].
The phrase Hermitian is also dropped out from self-(Hermitian)-adjoint
and it is taken as granted in Hermitian quantum mechanics.
Nevertheless, while investigating the real spectrum of non-Hermitian
Hamiltonians, we have to dis-entangle these two.
Apparently, the adjoint transformation brings about a ``trivial'' change
in case of differential operator, however, for a matrix operator it changes
rows to columns, which appears to be quite  a``non-trivial'' action. Notice that,
in matrix notation, we have
\begin{equation}
(Au)^\prime.v-u^\prime.A^\prime v=0,
\end{equation}
iff $^\prime$ denotes the transpose of a matrix and dot denotes matrix multiplication.
In matrix algebra, incidentally one defines ``adjoint'' of a matrix as $Adj(A)=A^{-1} |A|$,
which should be taken as a misnomer for quantum mechanical discussions.
Let us keep in mind that  $(p_x)^\ast=-p_x$ and the following transformations
\begin{equation}
T p_x T^{-1}=-p_x=P p_x P^{-1},T x T^{-1}=x=P (-x) P^{-1},
T KT^{-1}=K=P K P^{-1},
\end{equation}
for further discussions.
\par We propose to call a Hamiltonian, $H$, as pseudo-real if
\begin{equation}
\brho H\brho^{-1} =H^\ast,
\end{equation}
and pseudo-adjoint if
\begin{equation}
\bmu H\bmu^{-1} =H^\prime.
\end{equation}
{\bf Proposition I :}\\
If a Hamiltonian, $H$, is pseudo-real (10), then it has real eigenvalues, $E$,
subject to a condition on its eigenstate, $\Psi$.
Recall that $(A B)^\ast= A^\ast B^\ast$.\\
{\bf Proof :} Let $H \Psi=E \Psi$,
\begin{equation}
\Rightarrow (H \Psi)^\ast =(E\Psi)^\ast, \Rightarrow H^\ast \Psi^\ast
= E^\ast \Psi^\ast, \Rightarrow \brho H \brho^{-1} \Psi^\ast =E^\ast \Psi^\ast,
\Rightarrow H\brho^{-1}\Psi^\ast =E^\ast \brho^{-1} \Psi^\ast.\Box
\end{equation}
We finally find that
\begin{equation}
E=E^\ast,~~ \mbox {iff}~~ \brho^{-1} \Psi ^\ast = \epsilon \Psi.
\end{equation}
Let us have a quick illustration of what we mean. if $H_0=cp_x$, we find that
this Hamiltonian is pseudo-real under parity $P$, it possesses real eigenvalues
$\pm ck$ and the eigenstates are $\Psi=e^{\pm ikx}$, with $\epsilon=1.$\\
{\bf Proposition II : }\\
If a Hamiltonian, $H$, is pseudo-real (10) and pseudo-adjoint (11), then it
is pseudo-Hermitian (1) under
\begin{equation}
\bta=(\bmu\brho^{-1})^\prime.
\end{equation}
Recall that $(A B)^\prime=B^\prime A^\prime$. \\
{\bf Proof :}
\begin{equation}
\brho H \brho^{-1}=\brho (\bmu^{-1} H^\prime \bmu) \brho^{-1},\Rightarrow H^\ast=
\left ( \brho \bmu^{-1} H^\prime \bmu \brho^{-1} \right), \Rightarrow H^{\ast\prime}
=(\bmu \brho^{-1})^\prime~H~ (\brho \bmu^{-1})^\prime. ~~~~~~\Box
\end{equation}
Finally we have
\begin{equation}
(\bmu \brho^{-1})^\prime~ H~ \left ((\bmu \brho^{-1}) ^\prime\right )^{-1}=H^\dagger.
\end{equation}
Further, the orthogonality of the eigenstates will follow according to (2),
which now reads as
\begin{equation}
(E_m^\ast-E_n)~\Psi^\dagger_m ~(\bmu \brho^{-1})^\prime~ \Psi_n =0,~~~N_{\bta,n}
=\Psi^{\dagger}_n ~(\bmu \brho^{-1})^\prime~ \Psi_n.
\end{equation}
Hermiticity of $H$, follows when we have $\brho=\bmu$.  PT-symmetry
of the Hamiltonian follows when we have $\brho=P$ and $\bmu={\bf 1}$. In addition to
this, if we treat complex conjugation as $T$ in (13), we
re-discover the fact that eigenvalues of a PT-symmetric potential will be real
provided $PT\Psi =\epsilon \Psi$, i.e $\Psi$ is also the eigenstate of $PT$.
The Hamiltonians of the type $H_1={p_x}^2/(2m)+V_e(x)+iV_o(x)$ [1-6], where $e$ and
$o$ denote even and odd functions are such examples. For such PT-symmetric potentials,
the self-adjointness of $H$ is implied $\bmu={\bf 1}$, and the following
orthogonality condition
\begin{equation}
(E_m-E_n) \Psi^\prime_m \Psi_n=0
\end{equation}
will also work, automatically. Notice the absence of $^\dagger$ in (18).
One can check that $H_1$ possesses real eigenvalues since it is pseudo-real,
$P H_1 P^{-1}=H_1^\ast$ and the condition (13) is explicitly satisfied
by the energy-eigenstates.
Several, exactly solvable models of PT-symmetric potentials [1-6] are available
for a verification.
\par The complex quasi-exactly solvable Hamiltonian [3]
\begin{equation}
H_2={p_x^2 \over 2m}-(z\cosh 2x -3 i)^2
\end{equation}
has first three eigenvalues (real if $z^2 \le 1/4)$ and eigenfunctions
known analytically. $H_2$ was termed as PT-symmetric under $T : i \rightarrow -i$,
and $P : x\rightarrow i\pi/2-x$. Notice that both the operations
do not commute [5]. We find that $H_2$ more appropriately is pseudo-real under the
transformation $\brho : x \rightarrow (i\pi/2-x)$ and self-adjoint ($\bmu={\bf 1}$).
The eigenfunctions [3] can be checked to satisfy the proposed condition (13).
\par Let us consider the following Hamiltonian
\begin{equation}
H_3={[p_x+i\beta x]^2 \over 2m}+{1 \over 2}m \alpha^2 x^2,
\end{equation}
which admits real eigenvalues and real eigenvectors [10].
We find that $H_3$ is trivially pseudo-real (10) under $\brho={\bf 1}$ and we will have
real eigenvalues and real eigenfunctions too [10]. Next, $H_2$ is
pseudo-adjoint (11) as $e^{-\beta x^2} H_3 e^{\beta x^2}=H^\prime_3.$
So we have $\bmu=e^{-\beta x^2}=\bta.$ Alternatively, we may take $H_2$ to
be pseudo-real under $\brho=P$ and then $\bta=e^{-\beta x^2} P,$ also see [15].
Obviously, in both the cases $H_3$ would rather be categorized as pseudo-
Hermitian despite being $PT$-symmetric.
\par Next let us consider the Hermitian Hamiltonian
\begin{equation}
H_4={[p_x-3\gamma x^2]^2 \over (2m)}+{1 \over 2}m\alpha^2 x^2.
\end{equation}
which has real eigenvalues.
One can readily check that $\brho=\bmu=P$, this leads to Hermiticity.
We find that $e^{-2i\gamma x^3} H_4 e^{2i\gamma x^3}=H_4^\ast=H_4^\prime$
that means we again have the situation of Hermiticity where
$\brho=\bmu=e^{-2i\gamma x^3}.$ Other interesting options are to choose
$\brho=P$ and $\bmu=e^{-2i\gamma x^3}$ or $\brho=e^{-2i\gamma x^2}$ and $\bmu=P$.
In both situations, we have $\bta=e^{-2i\gamma x^3}P$, see also [14].
]The $\bta$-norm (2) will be indefinite, $(-1)^n, n=0,1,2..$ .
\par Intriguingly, when we choose to see even a Hermitian Hamiltonian
(e.g. (21)) as PT-symmetric or pseudo-Hermitian both the norms are indefinite
(positive-negative). However, the Hermitian norm , namely, $\Psi^\dagger \Psi$
remains definite (positive). This point has earlier
been revealed and remarked [22,24], however, it is often overlooked (see e.g. [23],[26]).
\par Complex Morse potential $V^{C-M}(x)=(A+iB)^2e^{-2x}-(2c+1)(A+iB)e^{-x}$
which is non-PT-symmetric was found [5] to have real eigenvalues.
Notice that the real Morse potential is written as $V^{R-M}(x)=D^2 e^{-2x}
-(2C+1)D e^{-x}$ and $V^{C-M}(x)$ is nothing but $V^{R-M}(x-ia).$ The Hamiltonian
with this potential has been investigated [9] to be pseudo-Hermitian under $\bta=
e^{-2 a p_x}$. If real potentials $V(x)$ admit real eigenvalues then the potentials
$V(x-ia)$ are also found to possess identical eigenvalues. When real and imaginary parts of
$V(x-ia)$ are separated out, the re-written potential would actually appear to be ``different''
and even ``unrelated'' with $V(x)$. The equivalence of two spectra will be due to
the fact that the Hamiltonian $H(x)=p_x^2/(2m)+V(x)$ follows : $e^{-a p_x}
H(x-ia) e^{a p_x} = H(x)$. We find that $e^{-2ap_x} H(x-ia) e^{2ap_x}= H(x+ia)$
implying that $\brho=e^{-2ap_x}$ and $\bmu={\bf 1}$. Thus, both the orthogonality
conditions (17) and (18) will be satisfied. We have indefinite norms : $N_{PT,n}
=(-1)^n=N_{\bta,n}.$
\par Norm of the eigenstates is required to be positive definite for a
probabilistic interpretation of quantum mechanics. In this regard the existence
of $\bta_+$ in the form $(DD^\dagger)^{-1}$ [24] for a pseudo-Hermitian Hamiltonian
possessing real eigenvalues is very important. Currently, the indefiniteness
of pseudo-norms is proposed to indicate the presence of a Hidden symmetry, $C$ [19],
which mimics charge-conjugation symmetry ${\cal C}$ [20]. It has also been
proposed that it is the $CPT-norm$ that will be positive definite.
Consequently, the Hermitian Hamiltonians are $P-,T-,PT-,$ and $CPT-$ invariant [22] and
pseudo-Hermitian Hamiltonians are $C-,PT-,$ and $CPT-$ invariant [24].
$PT-$ and $CPT-$ norms are indefinite and definite respectively.
In these works [19,21,22,24] one is actually talking about generalized discrete symmetry
operators : $C,P,$ and $T$ [23].
\par Recently, $2 \times 2$ pseudo-Hermitian matrix Hamiltonians [16] have been found to
give rise to a certain novelties in the random matrix theory. In this theory, to study
fluctuation properties of energy-spectrum hitherto one has modeled Hamiltonians as
real-symmetric or Hermitian matrices. More recently such simple $2 \times 2$
matrix Hamiltonians are being found handy in bringing out interesting features
of PT-symmetry [25,26].
\par In the following, we take up examples of simple pseudo-Hermitian matrices,
for further demonstration of the pseudo-reality and pseudo-adjointness of
Hamiltonians.
\begin{eqnarray}
H_5 =\left [\begin {array} {cc} a+ib & c \\ c & a-ib
\end {array} \right],~~H_6 =\left [\begin {array} {cc} a+c  & ib \\  ib & a-c
\end {array} \right],~~H_7 =\left [\begin {array} {cc} a   & i(b-c) \\ i(b+c) & a
\end {array} \right];~ c^2 > b^2
\end{eqnarray}
The eigenvalues of these matrices are $a\pm \sqrt{c^2-b^2}$. In the following, we
make an interesting use of Pauli matrices.
For $H_5$, we find that $\brho=\sigma_x, \bmu={\bf 1}$, so $H_5$ is pseudo-Hermitian
under $\bta=\sigma_x$. One can check that $H_6$ is pseudo-real under $\brho=\sigma_z$
and $H_6=H_6^\prime$, so it is pseudo-Hermitian under $\bta=\sigma_z$ as we have
$\bmu={\bf 1}$ again. The Hamiltonian $H_7$ is pseudo-adjoint under $\sigma_x$
and it is pseudo-real under $\sigma_z$ to display pseudo-Hermiticity under
$\bta=\sigma_y$.\\
Let us define a real diagonal matrix ${\cal E}=diag[E_1,E_2,E_3,...,E_n]$, i.e.,
${\cal E}^\ast = {\cal E}$ and  ${\cal E}^\prime = {\cal E}$
{\bf Proposition III :}\\
If a complex Hamiltonian, $H$, possessing real spectrum is diagonalizable by
an operator $D$, it is pseudo-real (10) under $\brho= D^\ast D^{-1}$(converse is also
true). \\
{\bf Proof :}
\begin{equation}
D^{-1} H D = {\cal E}, \Rightarrow D^{-1~\ast} H^\ast D^\ast
={\cal E}^\ast, \Rightarrow D^{-1~\ast} \brho H \brho^{-1} D^\ast={\cal E}
\Rightarrow \brho=D^\ast D^{-1}. ~~~~~~~~~~~~~\Box
\end{equation}
Note an interesting property of $\brho$ namely $\brho \brho^\ast=1.$ \\
{\bf Proposition IV :}\\
If a Hamiltonian is diagonalizable by an operator $D$, it is pseudo-adjoint (11)
under $\bmu=(DD^\prime)^{-1}.$ \\
{\bf Proof :}
\begin{equation}
D^{-1} H D= {\cal E}, \Rightarrow  D^\prime H^\prime D^{-1~\prime}=
{\cal E}^\prime, \Rightarrow D^\prime \bmu H \bmu^{-1} D^{-1~\prime}={\cal E}
\Rightarrow \bmu=(DD^\prime)^{-1}.~~~~~~~ \Box
\end{equation}
{\bf Proposition V :}\\
If a Hamiltonian $H$ possessing real spectrum is pseudo-real under
$\brho=D^\ast D^{-1}$ and pseudo-adjoint under $\bmu=(D D^\prime)^{-1}$,
it is pseudo Hermitian under $\bta=(D D^\dagger)^{-1}$ (Converse is also true).
\par The proof follows straight from Proposition {\bf II}. When $H$ is Hermitian
$D$ will be unitary ($U^\dagger=U^{-1}$). We find that $\brho=U^\ast U^\dagger=\bmu$
and $\bta={\bf 1}$. Note that $\bmu$ is self-adjoint i.e., $\bmu=\bmu^\prime$.\\
{\bf Illustration :}\\
The following Hamiltonian $H_8$
\begin{eqnarray}
H_8 =\left [\begin {array} {cc} a+ib & c+id \\ c-id & a-ib
\end {array} \right],~~\Psi_1 =\left [\begin {array} {c} -e^{-i\theta} \\ e^{-i\phi}
\end {array} \right],~~\Psi_2 =\left [\begin {array} {c} e^{i\theta} \\ e^{-i\phi}
\end {array} \right],~~\Phi_1 =\left [\begin {array} {c} 1 \\ 0
\end {array} \right],~~\Phi_2 =\left [\begin {array} {c} 0 \\ 1
\end {array} \right]
\end{eqnarray}
is pseudo-real under $\sigma_x$ and possesses real eigenvalues $a \mp e$,
where $e=\sqrt{c^2+d^2-b^2}$ if $c^2+d^2 > b^2$. Here $\Psi_n$ are
eigenvectors of  $H$ and $\Phi_n$ provides a fundamental orthonormal basis.
$D$ can be constructed as $D=\sum_{n} \Psi_n \Phi_n^\prime.$ We find the expressions
for $\brho, \bmu$ and $\bta_+$ are
\begin{eqnarray}
\brho =\left [\begin {array} {cc} 1 & -2ie^{i\phi} \sin \theta \\ 0 &
e^{2i \phi} \end {array} \right],
~\bmu={\mbox{sec}^2 \theta \over 2} \left [\begin {array} {cc} 1
& -i e^{i \phi} \sin\theta \\ -i \sin \theta e^{i \phi}& \cos 2 \theta e^{2i\phi}\end{array} \right]
,~\bta_+ ={\mbox{sec}^2\theta \over 2}\left [\begin {array} {cc}
1   & -i\sin\theta e^{i\phi} \\ i \sin\theta e^{-i\phi} & 1 \end {array} \right].
\end{eqnarray}
We have introduced $\theta=\tan^{-1}(b/e)$ and $\phi=\tan^{-1}(d/c)$.
This illustration also displays the non-uniqueness of $\brho$. Using $\brho=\sigma_x$
and $\bmu$ as in (26), we can construct $\bta=(\bmu \sigma_x)^{\prime}.$
This metric $\bta$ will satisfy the orthogonality condition (2), however, it does
not yield the $\bta$-norm (2) of the vectors $\Psi_n$ as real,
whereas $\bta_+$-norm will be real and positive definite.
\par The PT-symmetric potentials in finite basis space yield finite dimensional
matrix Hamiltonians. In this regard, it is interesting to note that two-dimensional
and three-dimensional matrix Hamiltonians obtained [27] for the potentials of the type
$V(x)=ix^{2n+1}$ are pseudo-real where $\brho=\sigma_z$ and
\begin{eqnarray}
\brho =\left [\begin {array} {ccc} 1 & 0 & 0 \\ 0 & -1 & 0 \\ 0 & 0 & 1
\end {array} \right],
\end{eqnarray}
respectively.
Some more interesting aspects of finite, D-dimensional, PT-symmetric Hamiltonians
have recently been discussed [25,26].
\par In the end, we conclude that Hamiltonians having real discrete spectrum are
first pseudo-real (10), further they could be Hermitian, PT-symmetric or pseudo-Hermitian.
The separation of adjointness of an operator from the Hermitian-adjointness
is something which is natural when one investigates real spectrum of
non-Hermitian Hamiltonians. Consequent to this, we find that the Hamiltonians
will have real spectrum if they are pseudo-real provided the eigenstates meet
the condition (13). Further, the proposed pseudo-adjointness (11) helps in
fixing the inner-product of the states. And this brings pseudo-reality to its
logical end, that is, $\bta$-pseudo-Hermiticity, however, not without enriching and
supplementing it with a relaxed {\it necessary} condition (10) and a crucial
axillary condition (13) on the eigenstates for real eigenvalues. We wish that
the simple examples presented here would help in further extensions by providing
a deeper insight in to this subject.
\section*{References }
\begin{enumerate}
\item C.M. Bender and S. Boettcher, Phys. Rev. Lett. {\bf 80} (1998) 5243,\\
C. M. Bender, S. Boettcher, P. N. Meisinger, J. Math. Phys. {\bf 40} (1999) 2201.
C. M. Bender, G. V. Dunne, P. N. Meisinger, Phys. Lett. A {\bf 252} (1999) 253.
\item F. Cannata, G. Junker, J. Trost, Phys. Lett. A {\bf 246} (1998) 219.\\
M. Znojil, Phys. Lett. A {\bf 259} (1999) 220.\\
G. Levai and M. Znojil, J. Phys. A : Math. Gen. {\bf 33} (2000) 7165.\\
Z. Ahmed, Phys. Lett. A : {\bf 287} (2001) 295; {\bf 286} (2001) 231.\\
P. Dorey, C. Dunning, R. Taeto, J. Phys. : Math. Gen. A {\bf 34} (2001) 5679. \\
R. S. Kaushal and Parthasarthi, J. Phys. A : Math. Gen. {\bf 35} (2002) 8743. \\
C.-S.Jia, X.-L.Zeng,L.-T. Sun, Phys. Lett. A {\bf 300} (2002) 78.
\item  A. Khare and B.P. Mandal, Phys. Lett. A {\bf 272} (2000) 53.
\item  B. Bagchi and C. Quesne, Phys. Lett. A {\bf 273} (2000) 285.
\item  B. Bagchi, S. Mallik, C. Quesne, R. Roychoudhury, Phys. Lett. A {\bf 289} (2001) 34.
\item Z. Ahmed, Phys. Lett. A : {\bf 282} (2001) 343. \\
 Z. Ahmed, `A generalization for the eigenstates of complex PT-invariant
potentials with real discrete eigenvalues' (unpublished) (2001). \\
 B. Bagchi, C. Quesne and M. Znojil, Mod. Phys. Lett. A{\bf 16} (2001) 2047.\\
 G.S. Japaridze, J. Phys. A : Math. Gen. {\bf 35} (2002) 1709.
\item
 R. Nevanlinna, Ann. Ac. Sci. Fenn. {\bf 1} (1952) 108; {\bf 163} (1954) 222.\\
 L.K. Pandit, Nouvo Cimento (supplimento) {\bf 11} (1959) 157.\\
 E.C.G. Sudarshan, Phys. Rev. {\bf 123} (1961) 2183.\\
 M.C. Pease III, {\it Methods of matrix algebra} (Academic Press, New York, 1965)\\
 T.D. Lee and G.C. Wick, Nucl. Phys. B {\bf 9} (1969) 209.\\
 F.G. Scholtz, H. B. Geyer and F.J.H. Hahne, Ann. Phys. {\bf 213} (1992) 74.
\item A. Mostafazadeh, J. Math. Phys. {\bf 43} (2002) 3944.
\item Z. Ahmed, Phys. Lett. A {\bf 290} (2001) 19.
\item Z. Ahmed, Phys. Lett. A {\bf 294} (2002) 287.
\item L. Solombrino, J. Math. Phys. {\bf 43} (2002) 5439.
\item G.Solarici, J. Phys. A: Math. Gen. {\bf 35} (2002) 7493.
\item T.V. Fityo, J. Phys. A : Math. Gen. {\bf 35} (2002) 5893.
\item A. Mostafazadeh, Mod. Phys. Lett. A {\bf 17} (2002) 1973.
\item B. Bagchi and C. Quesne, Phys. Lett. A {\bf 301} (2002) 173.
\item Z. Ahmed and S.R. Jain, ``Pseudo-unitary symmetry and the Gaussian
pseudo-unitary ensemble of random matrices'', arXiv quant-ph/0209165,
to appear in Phys. Rev. E)\\
Z. Ahmed and S.R. Jain, J. Phys. A: Math. Gen. {\bf 36} (2003) 3349
(The special issue on Random Matrices). \\
Z. Ahmed, Phys. Lett. A {\bf 308} (2003) 140.
\item e.g., P. M. Morse and H. Feshbach, {\em Methods of Theoretical Physics}
vol. I (Mc Graw-Hill Book Company, Inc., New York, 1953) 866-886.
(On page 884, one may see a discussion on non-Hermitian operators
and Biorthogonal functions.)
\item e.g., J. Von. Neumann, {\em Mathematical Foundations of Quantum Mechanics}
(Princeton University Press, Princeton, 1955). \\
E. Merzbacher, {\em Quantum Mechanics} (John Wiley and Sons, Inc, New York,
1970).
\item C.M. Bender, D.C.Brody and H.F.Jones, Phys. Rev. Lett. {\bf 89} (2002)
270401.
\item T. D. Lee, {\em Particle Physics and Introduction to Field Theory}, vol. 1,
(Harwood academic puplishers, New York, 1981).\\
J.D. Bjorken and S. D. Drell, {\em Relativistic Quantum Fields} (McGraw-Hill Inc., New York, 1965).
\item C. M. Bender, P. N. Meisinger and Q. Wang, J. Phys. A : Math. Gen. {\bf 36}
(2003) 1029.
\item Z. Ahmed, Phys. Lett. A  {\bf 310} (2003) 139.
\item A. Mostafazadeh, J. Math, Phys. {\bf 44} (2003) 974.
\item Z. Ahmed, `C-,PT-, and CPT- invariance of pseudo-Hermitian Hamiltonians',
arXiv quant-ph/0302141.
\item C. M. Bender, P.N. Meisinger, Q. Wang, `Finte-Dimensional PT-symmetric Hamiltonians'
arXiv quant-ph/0303174.
\item A. Mostafazadeh, `Exact PT-symmetry is equivalent to Hermiticity', arXiv
quant-ph/0304080.
\item C.K.Mondal, Kaushik Maji, S. P. Bhatacharyya, Phys. Lett. A {\bf 291} (2001) 203.
\end{enumerate}
\end{document}